\documentclass[11pt,letterpaper]{article}
\usepackage[letterpaper,margin=1in]{geometry}

\usepackage[margin=1in]{geometry}
\usepackage{graphicx}
\usepackage{amsmath}
\usepackage{booktabs}
\usepackage{url}
\usepackage{multirow}
\usepackage{color}	
\usepackage{soul}
\usepackage{algorithm}
\usepackage{algorithmic}
\usepackage{caption}
\usepackage{subcaption}
\usepackage{paralist}
\usepackage{amssymb}
\usepackage[skipabove=\topskip,skipbelow=\topskip,nobreak=true]{mdframed}
\usepackage[dvipsnames]{xcolor}
\usepackage[colorlinks=true, citecolor=blue, linkcolor=blue, urlcolor=blue]{hyperref}
\usepackage{authblk}

\hypersetup{
  pdftitle={VCG: A Multimodal Retrieval Framework for E-Commerce Video Feeds under Extreme Cold-Start Conditions},
  pdfauthor={Katya Mirylenka et al.},
  pdfsubject={Multimodal retrieval for e-commerce video feeds under extreme cold-start conditions},
  pdfkeywords={Recommender Systems, Multimodal Embeddings, Vector Search, Video Retrieval, E-Commerce, Cold-Start}
}

\title{\textbf{VCG: A Multimodal Retrieval Framework for E-Commerce Video Feeds under Extreme Cold-Start Conditions}}

\author[1]{Katya Mirylenka\thanks{Work done while at Zalando Switzerland AG.}}
\author[2]{Egor Malykh}
\author[2]{Mahdyar Ravanbakhsh}
\author[3]{Michael Gygli}
\author[3]{Marco-Andrea Buchmann}
\author[2]{Andrew Dzhoha}
\author[2]{Svitlana Borzenko}
\author[3]{Francesca Catino}
\author[2]{Mohamed Gaafar}
\author[2]{Maarten Versteegh}
\author[2]{Thomas Kober}
\author[3]{Dario d'Andrea}
\author[2]{Ellie Langhans}

\affil[1]{TU Wien, Vienna, Austria \authorcr \texttt{katsiaryna.mirylenka@tuwien.ac.at}}
\affil[2]{Zalando SE, Berlin, Germany \authorcr \texttt{first\_name.last\_name@zalando.de}}
\affil[3]{Zalando Switzerland AG, Zurich, Switzerland \authorcr \texttt{first\_name.last\_name@zalando.ch}}

\date{}

\begin{document}

\maketitle

\begin{abstract}
The digital commerce landscape is shifting from static, search-driven catalogs to dynamic, immersive video feeds. This transition introduces an ``extreme cold-start'' problem: unlike traditional items, new short-form videos lack the dense interaction history required for collaborative filtering. Furthermore, immersive feeds introduce strong position and duration biases that distort standard engagement signals. In this paper, we demonstrate the Video Candidate Generation (VCG) system, a scalable multimodal retrieval engine designed to solve these challenges in a large-scale e-commerce environment. By leveraging a domain-adapted vision-language model (based on CLIP), we map users and videos into a shared semantic space, enabling zero-shot retrieval based on visual content rather than behavioral history. We detail the system's architecture and present a rigorous evaluation comparing generative (LLM) vs. discriminative (CLIP) embeddings. Our results show that while generative models excel at attribution, they suffer from embedding space collapse in retrieval tasks. Online A/B testing demonstrates that VCG effectively mitigates engagement biases, yielding a 50\% uplift in deep video completion. To showcase the system's capabilities, we present an interactive demonstration featuring three bi-directional retrieval scenarios: Product-to-Video, Video-to-Product, and Zero-Shot Semantic Search.

\end{abstract}

\section{Introduction}
Traditionally, e-commerce platforms have operated as digital catalogs where the user journey is goal-oriented: a user enters with specific intent, executes a search, and transacts. 
In a catalog environment, products are long-lived and accumulate ample interaction data, which naturally enables robust representation learning for downstream tasks like product recommendation and customer modeling \cite{mirylenka2019hidden}.
However, the success of social media platforms has conditioned users to expect implicit discovery via infinite feeds. To align with these expectations, we have introduced a video feed on a large fashion e-commerce website.

Transitioning to a feed-based architecture introduces an Extreme Cold-Start problem. In a catalog, products are long-lived and accumulate ample interaction data. In a video feed, content is ephemeral and voluminous; new videos have no interaction history, rendering standard Matrix Factorization or Autoencoder-based recommenders ineffective. Furthermore, optimizing for standard ``watch-time'' introduces duration bias, favoring shorter videos regardless of relevance \cite{zhan2022deconfounding}, as well as position bias (favoring earlier content).

To address this, we propose a \emph{semantic-first} recommendation approach. Instead of asking \emph{Who else watched this video?} (collaborative signal), our Video Candidate Generation (VCG) system asks \emph{What does this video look like, and does it match the user's visual style?} (semantic signal). 

\textbf{Our Contributions are the following:}
\begin{compactitem}
    \item \textit{Zero-Shot Two-Tower Architecture}: We demonstrate a retrieval system that decouples representation from interaction. By using pre-trained multimodal encoders, we bypass the need for massive training datasets.
    \item \textit{Generative Adjudication Protocol}: We introduce an ``LVLM-as-a-judge'' evaluation framework. We show that while traditional offline metrics fail to predict online success due to exposure bias, a Large Vision-Language Model (Qwen-VL) successfully acts as a proxy for human relevance.
    \item \textit{Embedding Space Analysis}: We provide a comparative analysis of generative vs. discriminative embeddings for retrieval, demonstrating why contrastive models (CLIP) outperform generative models (LLMs) for vector search tasks in e-commerce.
\end{compactitem}

\section{Background and Related Work} \label{sec:rel_work}

\textbf{Two-Tower Architectures:} Our work builds upon the dual-encoder paradigm popularized by YouTube \cite{covington2016deep} and Pinterest \cite{pal2020pinnersage}. While foundational work in this space focuses on correcting sampling bias in large corpora \cite{yi2019sampling}, these systems primarily train encoders on interaction data. VCG, in contrast, uses pre-trained multimodal encoders to function in zero-shot scenarios where interaction graphs are sparse. This approach structurally parallels work in data integration, where Siamese architectures and graph neural networks have been successfully deployed for entity matching across disparate sources without requiring explicit interaction bridges~\cite{krivosheev2021business, krivosheev2023graph}.

\textbf{Multimodal Learning:} The core engine of VCG is a domain-adapted version of CLIP \cite{radford2021learning}. Similar approaches have been applied to domain-specific tasks, such as FashionCLIP \cite{chia2022contrastive}, which aligns fashion imagery with textual attributes. Adapting representations to highly specific, low-resource e-commerce domains is a well-documented challenge; for instance, acquiring label-efficient representations for domain-specific text categorization traditionally requires complex reinforced active learning pipelines \cite{wertz2022evaluating, wertz2022investigating, wertz2023reinforced}. By transitioning directly to a zero-shot multimodal space, VCG bypasses the need for such costly human-in-the-loop metadata annotation entirely.

\textbf{Generative vs. Discriminative Models:} 
Recent research has broadly explored leveraging Large Language Models (LLMs) to bridge semantic gaps in structured data retrieval, ranging from zero-shot rankers \cite{hou2024large} to systems that adapt LLMs for seamless natural language API integration and robust Text-to-SQL conversion \cite{chan2024adapting, dragusin2025grounding, hampp2026optimizing}. However, while these generative models excel at translating abstract user intent into structured logical constraints and metadata queries, applying them directly to generate embeddings for raw visual retrieval tasks often leads to \textit{anisotropy} and representation collapse—a phenomenon where embeddings occupy a narrow cone in the vector space, severely limiting discriminative power \cite{ethayarajh2019contextual}.

\textbf{Evaluation with LLMs:} Addressing the exposure bias in offline metrics requires novel evaluation strategies. 
We adopt the ``LLM-as-a-judge'' paradigm \cite{zheng2024judging,hosseini2024llmasajudge}, utilizing large models to simulate human preference and provide a scalable alternative to manual annotation. While our current demonstration relies on constrained, direct prompt-based scoring, mitigating the inherent risks of hallucination and variance in generative evaluators remains critical for production systems. Future enhancements to our evaluation protocol could incorporate robust uncertainty quantification frameworks—such as testing the consistency hypothesis and applying similarity-based aggregation (SIMBA) \cite{xiao2025consistency, bhattacharjya2025simba}. Recognizing and quantifying model uncertainty as a safeguard aligns with our broader ongoing initiatives toward reliable conversational data analytics \cite{amer2025towards}.


\section{System Architecture and Methodology} \label{sec:method}
\begin{figure}[h]
  \centering
  \includegraphics[width=0.8\linewidth]{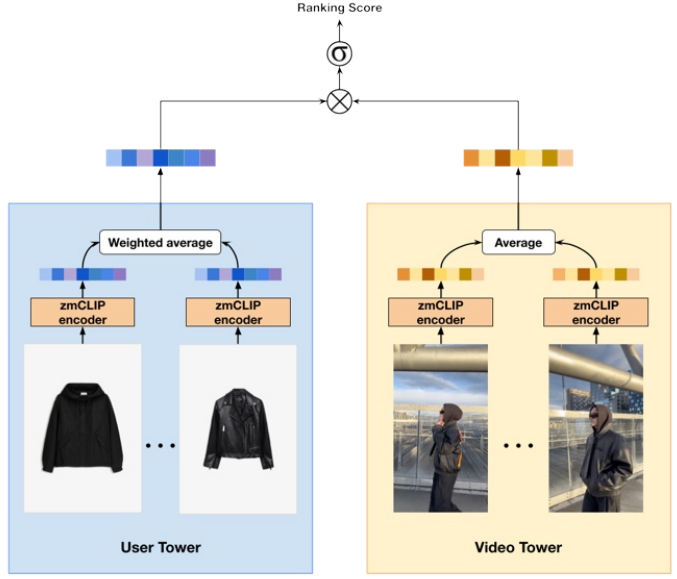} 
  \caption{VCG v2 Architecture: Video embeddings are pre-computed via frame aggregation (right), while User embeddings are computed dynamically from interaction history (left).}
  \label{fig:arch}
\end{figure}

To enhance the relevance and engagement of our video feed, we developed the \textbf{Video Candidate Generation (VCG)} system~\cite{dzhoha2025short}. This system utilizes a scalable Two-Tower architecture (Figure \ref{fig:arch}) designed for high-throughput inference, effectively decoupling interaction history.

\subsection{Evolution from VCG v1 (metadata-based)}
Prior to the multimodal approach presented here, we developed and tested an initial version, VCG v1. This iteration utilized a conventional supervised Two-Tower architecture. It reused the existing, highly optimized User Tower from our main catalog recommendation system to generate non-trainable user embeddings~\cite{celikik2024buildingscalableeffectivesteerable}. A new video tower was trained from scratch using available video metadata (creator IDs, hashtags, associated product links).

Despite being a theoretically ``stronger'' baseline than simple recency-based methods, VCG v1 failed to deliver uplift in online A/B tests. Our analysis identified three critical failure modes:
\begin{compactenum}
    \item \textbf{Semantic gap:} User representations were trained on ``Add-to-Cart'' and purchase signals. This strong conversion intent did not align with the inspirational intent of the video feed.
    \item \textbf{Metadata sparsity:} Reliance on creator-generated tags resulted in poor coverage. The model defaulted to popularity bias.
    \item \textbf{Lack of visual signal:} The model ignored the actual pixel content of the videos, failing to capture aesthetic preferences (e.g., \emph{Boho style} vs. \emph{Streetwear}).
\end{compactenum}
These insights drove the pivot to VCG v2 (multimodal), which creates a shared embedding space for users and videos based on visual content.

\subsection{Multimodal Representation (VCG v2)}

\noindent \textbf{Video Representation:} A video $v \in V$ is treated as a ``bag-of-frames''. We extract $N=10$ frames uniformly sampled from the video. Each frame is encoded with our in-house CLIP-based model, fine-tuned using a large-scale fashion dataset comprising approximately 200 million image–text pairs. The final video embedding $e_v$ is the mean of these frame vectors:
\begin{equation}
    e_v = \frac{1}{N}\sum_{j=1}^{N}E_{\text{CLIP}}(\text{frame}_{j}).
\label{eq:emb}
\end{equation}
We explicitly exclude textual metadata to avoid the noise encountered in VCG v1.

\noindent \textbf{User Representation via Domain Adaptation:} Let $U$ be the set of users. For each user $u \in U$, we have a time-ordered history of catalog interactions $H_{u}=[(s_{1},t_{1}),...,(s_{n},t_{n})]$, where $s_{k}$ is a product and $t_{k}$ is the timestamp. 
Evaluating distance metrics across such time-ordered sequences is critical for establishing robust similarity measures, a foundational challenge that remains significant in uncertain data environments \cite{mirylenka2017data, dallachiesa2011similarity, dallachiesa2012uncertain}.
A user is represented by a dynamic composition of these interactions:
\begin{equation}
    e_{u} = \frac{\sum_{k=1}^{|H_u|} w_k \cdot E_{\text{CLIP}}(s_k)}{\sum_{k=1}^{|H_u|} w_k}, \quad w_{k} = \exp(-\lambda(t_{\text{now}}-t_{k})).
\end{equation}
Crucially, the embedding function $E_{\text{CLIP}}$ is a domain-adapted version. This process aligns the vector space such that ``visual'' video content and ``commercial'' product content reside in the same manifold.

\subsection{Problem Formulation and Inference}
Let $U$ be the set of users and $V$ be the set of creator videos. For each user $u \in U$, we have a time-ordered history of interactions with items from our product catalog, $H_{u}=[(p_{1},t_{1}),(p_{2},t_{2}),..., (p_{n},t_{n})],$ where $p_{i}$ is a product item and $t_{i}$ is the interaction timestamp. Our goal is to learn a scoring function $f: U \times V \rightarrow \mathbb{R}$ that predicts the relevance of a video $v$ to a user $u$. The final output for each user is a ranked list of videos such that highly relevant videos appear first.

The scoring function is modeled as the dot product similarity between user and video embeddings in a shared $d$-dimensional space derived from our domain-adapted CLIP model: $f(u,v) = {e_{u}}^{T}e_{v}$. Because video embeddings are pre-computed offline, online inference is restricted to a rapid vector similarity search, achieving a median latency (P50) of 17.5ms and a tail latency (P99) of 30ms.


\section{Experimental Setup and Results} \label{sec:experim}

\subsection{Evaluation Framework: The Paradox}
Standard offline metrics (NDCG, AUC) computed on historical logs showed no significant improvement over a recency baseline due to \textit{exposure bias}: historical logs are generated by the old system, unable to validate the relevance of new videos.

To resolve this, we adopted an \textbf{LVLM-as-a-judge} approach~\cite{zheng2024judging,hosseini2024llmasajudge}. We utilized Qwen 2.5-VL as an external evaluator. The prompt (Figure \ref{fig:llm_prompt}) acts as a ``system instruction'', enabling the model to rate the visual coherence between a user's history and a recommended video on a standard 5-point Likert scale.

\begin{figure}[h]
    \centering
    \begin{mdframed}[backgroundcolor=gray!10, linecolor=gray!50, linewidth=1pt, roundcorner=5pt, innertopmargin=10pt, innerbottommargin=10pt, innerrightmargin=10pt, innerleftmargin=10pt]
        \small 
        \textbf{System Instruction:} You are an AI fashion relevance analyst. Your primary function is to critically and objectively evaluate the relevance of video content against a specific user's fashion history. It is crucial that you use the defined textual relevance categories appropriately and avoid defaulting to a generally positive assessment unless there is substantial, specific evidence.
        
        \vspace{0.5em}
        $\langle...\rangle$ 
        
        \vspace{0.5em}
        Assign one of the following textual categories for relevance. Choose the category that most accurately describes the alignment. Be discerning.
        \begin{itemize}
            \setlength\itemsep{0.1em} 
            \item \texttt{"excellent\_match"}: $\langle...\rangle$
            \item \texttt{"good\_match"}: $\langle...\rangle$
            \item \texttt{"partial\_match"}: $\langle...\rangle$
            \item \texttt{"poor\_match"}: $\langle...\rangle$
            \item \texttt{"no\_match"}: $\langle...\rangle$
        \end{itemize}
    \end{mdframed}
    \caption{The instruction prompt provided to the Qwen-VL judge for evaluating visual coherence.}
    \label{fig:llm_prompt}
\end{figure}

\subsection{Attribute Prediction and Model Selection}
To rigorously justify our choice of CLIP over newer generative models for the retrieval core, we evaluated both model families on an auxiliary task: \textit{Zero-Shot Attribute Prediction}. 

We trained lightweight classifiers (Multilayer Perceptrons) on top of the raw video embeddings extracted from CLIP and Qwen2.5-Omni/VL (using the final hidden layer representations) to predict video metadata tags such as ``content theme'' (e.g., fashion, sports) and ``editorial format'' (e.g., unboxing, review).

As shown in Table \ref{tab:attribute_pred}, Qwen embeddings achieved a consistent 6-10\% uplift in F1-scores compared to CLIP. This indicates that generative models capture richer semantic nuances and are superior for offline content enrichment tasks.


\begin{table}[h]
\centering
\small
\caption{Zero-Shot Attribute Prediction (F1-Score). Qwen outperforms CLIP on classification tasks.}
\label{tab:attribute_pred}
\begin{tabular}{lcc}
\toprule
\textbf{Task (Label)} & \textbf{CLIP} & \textbf{Qwen2.5-Omni} \\
\midrule
Content Theme  & 0.92 & \textbf{0.94} \\
Editorial Format  & 0.75 & \textbf{0.78} \\
Video Source Classification & 0.86 & \textbf{0.93} \\
\bottomrule
\end{tabular}
\end{table}

While Qwen embeddings achieved a 6-10\% uplift in F1-scores, they exhibited high anisotropy during semantic retrieval (all vectors clustered in a narrow cone), leading to poor separability in k-NN search. CLIP's contrastive loss function enforces a more uniform distribution on the hypersphere, making it the superior choice for the retrieval engine. 

\subsection{Offline Evaluation: Visual Coherence}
We introduced a metric called \textbf{visual coherence}, measured by similarity in an alternative purely image-based embedding space~\cite{Bracher2016fdna}, alongside semantic similarity gains in the CLIP embedding space.

\begin{table}[h]
\centering
\small
\caption{Offline Evaluation: VCG vs. Recency Baseline. Note the high gains in semantic similarity metrics.}
\label{tab:offline}
\begin{tabular}{lcccc}
\toprule
\textbf{Method} & \textbf{fDNA S.} & \textbf{CLIP S.} & \textbf{LLM Top-5} & \textbf{LLM Top-10} \\
\midrule
Recency & 13.8 (6.48) & 0.41 (0.06) & 2.72 (0.30) & 2.43 (0.24) \\
VCG & \textbf{18.9 (8.23)} & \textbf{0.50 (0.04)} & \textbf{3.12 (0.35)} & \textbf{3.09 (0.32)} \\
\midrule
\textit{Gain} & \textit{+37\%} & \textit{+22\%} & \textit{+14.7\%} & \textit{+27\%} \\
\bottomrule
\end{tabular}
\end{table}

VCG improved all offline semantic-coherence metrics over the recency baseline (Table \ref{tab:offline}): a 37\% gain in fDNA~\cite{Bracher2016fdna} similarity and a 22\% gain in CLIP similarity. The LVLM judge corroborated this, significantly shifting the distribution towards ``Relevant'' (from 2.43 to 3.09 for Top-10). 

\begin{figure}[h]
  \centering
  \includegraphics[width=0.5\linewidth]{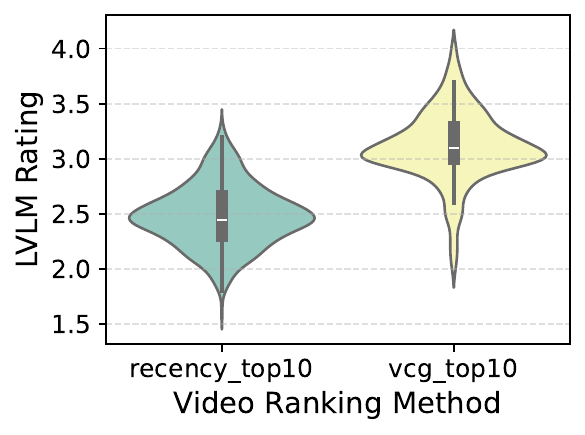} 
  \caption{Comparison of VCG with recency-based baseline using LVLM-as-a-judge scores. VCG consistently shifts the distribution towards higher relevance ratings.}
  \label{fig:violin}
\end{figure}

To visualize this qualitative shift, Figure \ref{fig:violin} presents the full distribution of the LVLM judge scores. While the recency baseline exhibits a distribution centered in the lower-to-mid range, the VCG model heavily skews the density towards ``good\_match'' and ``excellent\_match'' classifications, confirming the system's superior visual and semantic alignment.

\subsection{Online: A/B Testing}
A 4-week online A/B test (Table \ref{tab:online}) confirmed that semantic relevance drives engagement. 

\begin{table}[h]
\small
\centering
\caption{Online A/B Test Results (Treatment vs. Control)}
\label{tab:online}
\begin{tabular}{lcc}
\toprule
\textbf{Metric} & \textbf{Lift} & \textbf{Confidence Interval (95\%)} \\
\midrule
VideoProgress @ 25\% & \textbf{+40.97\%} & [21.25, 60.69]\% \\
VideoProgress @ 50\% & \textbf{+50.10\%} & [22.03, 78.17]\% \\
Video Start Rate & +8.17\% & [-0.57, 16.91]\% \\
Start $\rightarrow$ 25\% Conv. & \textbf{+30.32\%} & [16.69, 43.96]\% \\
Start $\rightarrow$ 50\% Conv. & \textbf{+38.76\%} & [17.01, 60.52]\% \\
\bottomrule
\end{tabular}
\end{table}


While the lift in \textit{video start rate} (+8\%) was modest, the gains in consumption depth were substantial (+50\% for videos watched at least halfway). This indicates that while users started slightly more videos, the videos they did start were far more relevant, keeping them engaged for longer. This effectively counters the clickbait problem where high click-through rates mask poor content fit. 

\begin{figure}[h]
  \centering
  \includegraphics[width=0.99\linewidth]{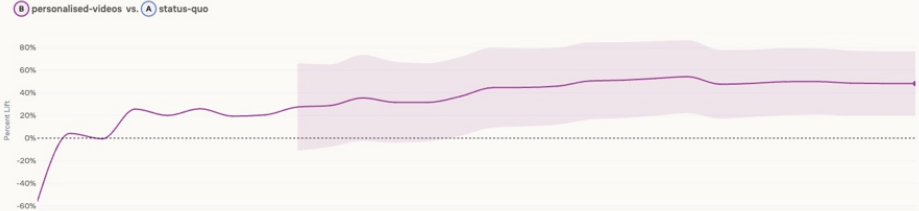} 
  \caption{Time series of Percent Lift for Video Progress @ 50\%. The VCG model consistently outperforms the baseline throughout the testing period.}
  \label{fig:lift}
\end{figure}

To illustrate the stability of this engagement increase over the duration of the experiment, Figure \ref{fig:lift} plots the daily percent lift for the \texttt{VideoProgress@50\%} metric. The time series demonstrates that the VCG model's outperformance is not an artifact of a short-term novelty effect, but a consistent, sustained improvement over the control baseline throughout the entire 4-week testing period. Importantly, core business metrics (revenue, user retention rate) remained stable, proving that the video feed adds value without cannibalizing transaction flow.

\section{Demonstration Scenarios and Interface} \label{sec:demo}
\begin{figure}[h]
  \centering
  \includegraphics[width=0.32\linewidth, height=3.4cm]{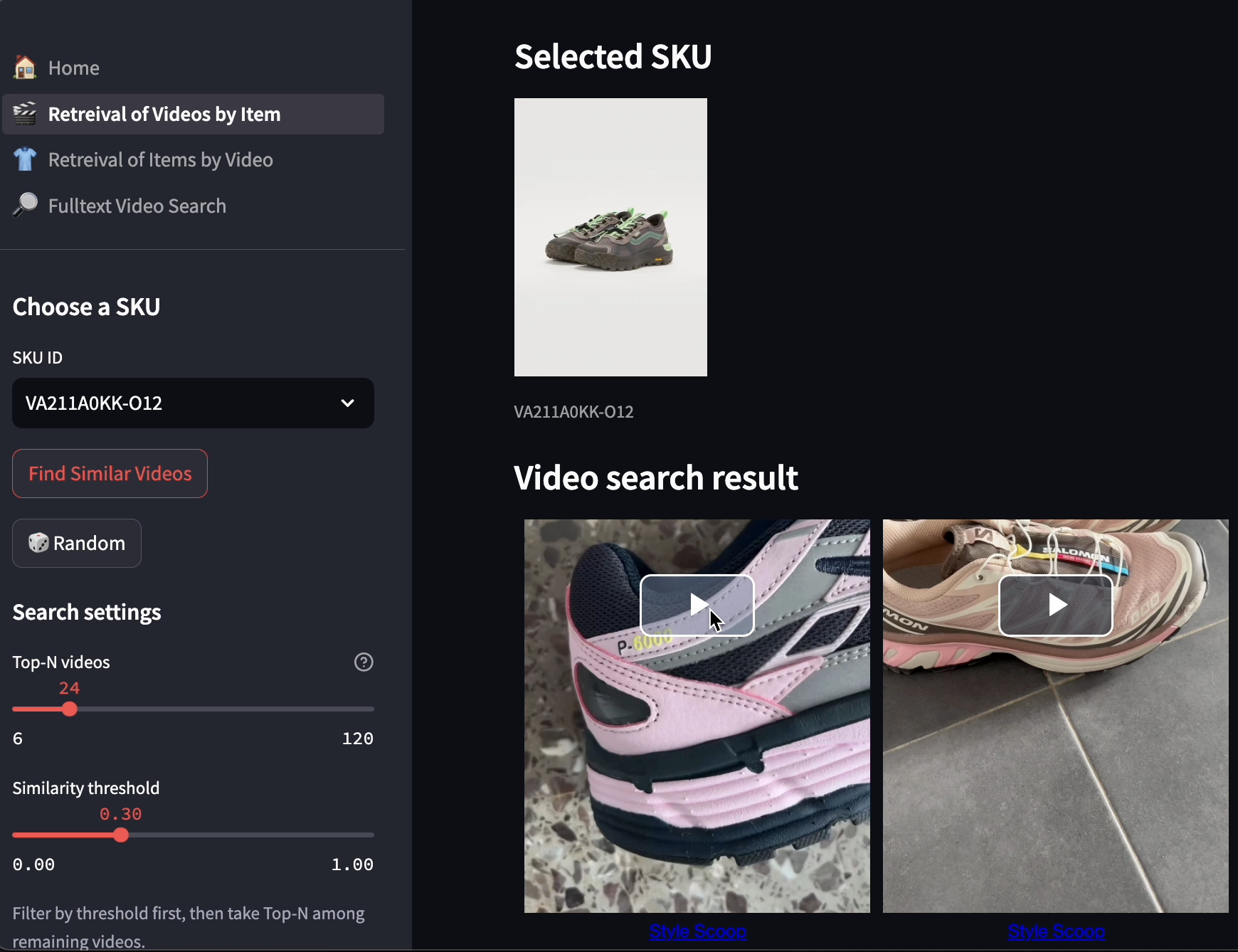}\hfill
  \includegraphics[width=0.32\linewidth, height=3.4cm]{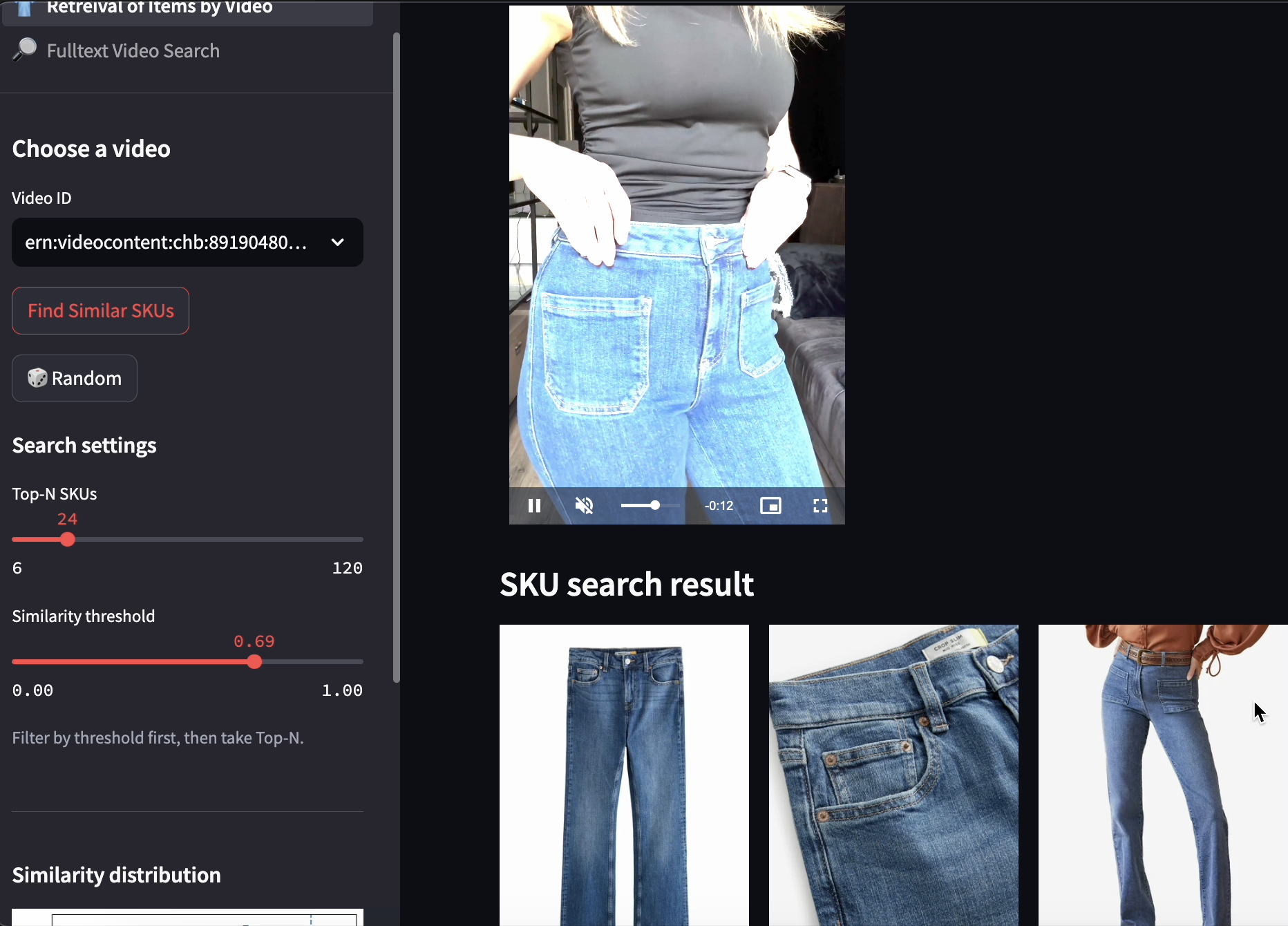}\hfill
  \includegraphics[width=0.32\linewidth, height=3.4cm]{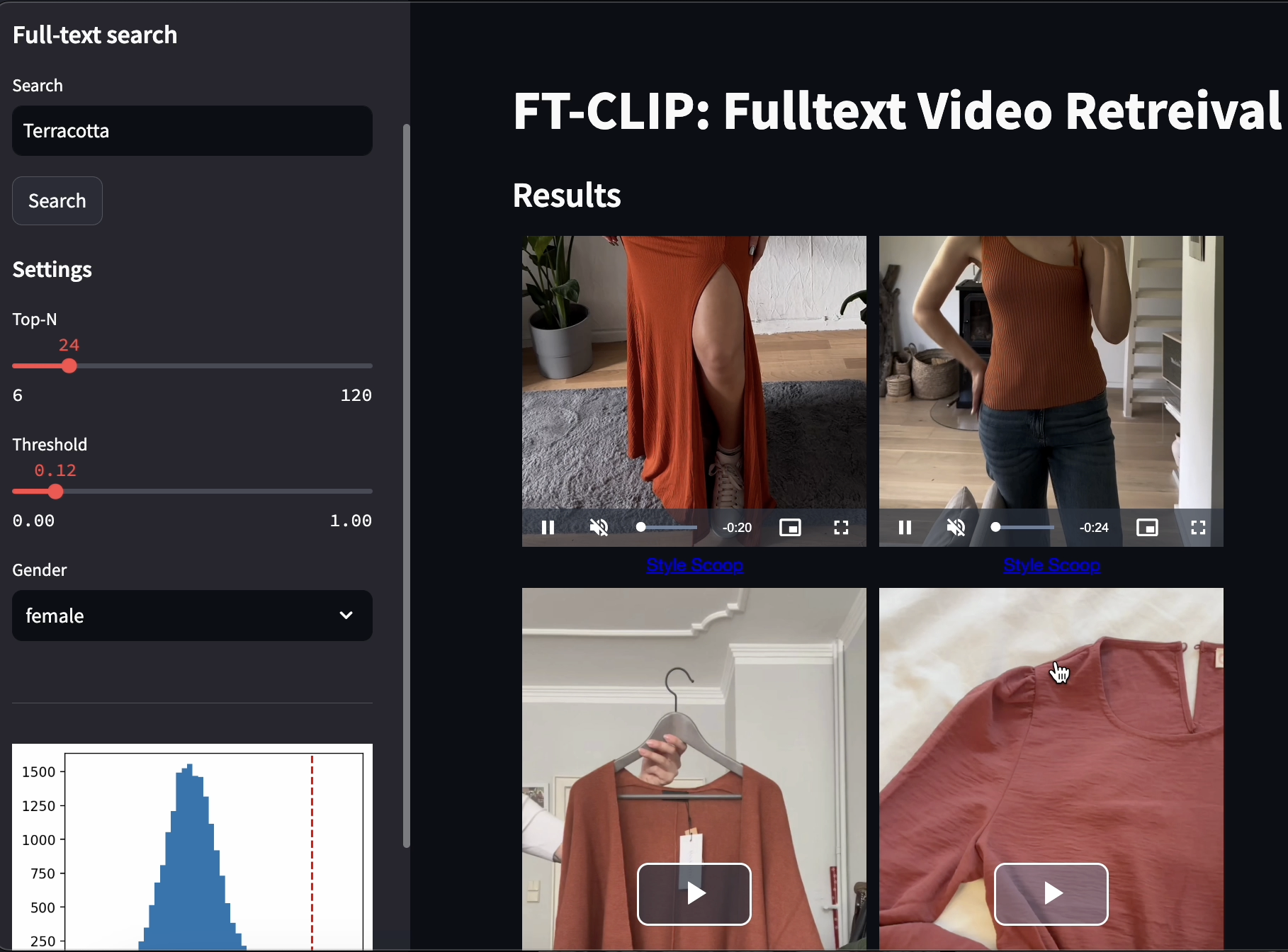}
  \caption{The VCG Demonstration Interface. \textbf{Left:} Scenario 1 (Product $\to$ Video) retrieving content for a selected sneaker. \textbf{Center:} Scenario 2 (Video $\to$ Product) retrieving matching denim SKUs for a selected video. \textbf{Right:} Scenario 3 (Text $\to$ Video) showing zero-shot semantic retrieval for the query ``Terracotta''.}
  \label{fig:ui}
\end{figure}

Our interactive demonstration\footnote{Demo video available at: \url{https://youtu.be/ClF6iv_PH4A}} showcases the versatility of the shared multimodal embedding space through a custom web UI (Figure \ref{fig:ui}). 

\begin{compactenum}
    \item \textbf{Get the Look (Product $\to$ Video):} The user selects a static product (e.g., a floral dress). The system uses the product embedding as a query vector to perform a k-NN search against the video index (Fig \ref{fig:ui}, Left), retrieving videos featuring similar items. This solves the \emph{content association} problem, automatically linking inventory to content without manual tagging.
    
    \item \textbf{Shop the Look (Video $\to$ Product):} The user selects a video. The system queries the product catalog using the video embedding (Fig \ref{fig:ui}, Center). A list of available products matching the visual style is displayed, enabling ``shoppable'' media from purely user-generated content.
    
    \item \textbf{Zero-Shot Semantic Search (Text $\to$ Video):} The user types a text query (e.g., ``Cyberpunk street style''). The text is encoded via the CLIP text encoder to query the video index (Fig \ref{fig:ui}, Right). The system retrieves relevant videos even if they lack text metadata/tags. This demonstrates the model's understanding of abstract visual concepts, vastly expanding content discoverability.
\end{compactenum}

\section{Discussion and Conclusion}
Standard e-commerce recommenders rely on collaborative filtering, requiring dense interaction histories. Off-the-shelf text search engines rely on metadata matching. As shown in Table \ref{tab:comparison}, VCG uniquely addresses the needs of ephemeral video by operating in a zero-shot, cross-modal capacity without relying on popularity proxies.

\begin{table}[h]
\centering
\caption{Comparison of VCG against standard industry methodologies.}
\label{tab:comparison}
\resizebox{0.9\textwidth}{!}{%
\begin{tabular}{@{}lccc@{}}
\toprule
\textbf{Capability} & \textbf{VCG (Ours)} & \textbf{Collaborative Filtering} & \textbf{Metadata Search} \\ \midrule
\textbf{Handles Extreme Cold-Start} & \textbf{Yes (Zero-Shot)} & No (Requires interactions) & Partial (Requires tags) \\
\textbf{Visual Semantic Understanding} & \textbf{High (Pixel-level)} & None & None \\
\textbf{Cross-Modal (Product $\leftrightarrow$ Video)} & \textbf{Yes (Shared space)} & No & No \\
\textbf{Susceptibility to Popularity Bias}& \textbf{Low} & High & High \\
\bottomrule
\end{tabular}%
}
\end{table}

\noindent \textbf{Responsible AI Aspects:} Decoupling retrieval from historical click-graphs may reduce dependence on popularity signals; measuring creator-level exposure remains future work. Because VCG maps users to videos based on objective aesthetics rather than creator identity or demographic tags, the system provides fairer exposure to long-tail and underrepresented creators.

\noindent \textbf{Conclusion:} The VCG system serves as a bridge between the structured, search-based past of e-commerce and the fluid, discovery-based future of video feeds. By decoupling representation from interaction history, we effectively addressed the extreme cold-start problem. The documented divergence between offline proxies (NDCG) and online reality serves as a crucial warning to the research community: in the era of generative AI, relevance is best measured by semantic coherence, not just historical clicks.

\section*{Acknowledgements}
We are deeply thankful for the contributions of our colleagues, Dmitry Isaev, Siobhan Hughes, and Varya Obolonchykova, for their invaluable help throughout this work. We also thank Tofigh Naghibi, Jacek Wasilewski, Ton Torres and Josip Krapac for their constant support, both technical and methodological.

\bibliographystyle{unsrt}
\bibliography{references_short}

\end{document}